\documentclass{article}

\usepackage[preprint]{tackling_climate_workshop_style}
\usepackage{amsmath,amssymb,amsthm}
\usepackage{booktabs}
\usepackage{float}
\usepackage{microtype}
\usepackage{pgfplots}
\usepackage{siunitx}
\usepackage{xcolor}
\usepackage{url}
\usetikzlibrary{arrows.meta,decorations.pathmorphing,positioning}
\pgfplotsset{compat=1.18}

\newtheorem{theorem}{Theorem}

\newtheorem{lemma}{Lemma}
\newtheorem{corollary}{Corollary}
\newcommand{\cE}{\mathcal E}
\newcommand{\cR}{\mathcal R}
\newcommand{\bone}{\mathbf 1}
\newcommand{\epscr}{\varepsilon_{\mathrm{cr}}}
\DeclareMathOperator{\tr}{tr}

\title{Congestion Structure and Exceedance Bounds for \\Locational Marginal Emissions}
\author{
Cameron Khanpour\\
Georgia Institute of Technology\\
\texttt{khanpour@gatech.edu}
\And
Samuel Talkington\\
Harvard University\\
University of Michigan\\
\texttt{talks@umich.edu}
\And
Daniel K. Molzahn\\
Georgia Institute of Technology\\
\texttt{molzahn@gatech.edu}
}

\begin{document}

\maketitle

\begin{abstract}
Locational marginal emissions (LMEs) give the sensitivity of total operating carbon
emissions to nodal power demand. We show that this vector with \(n\) entries has a
much smaller intrinsic dimension under DC optimal power flow. Within a fixed
active constraint set, the LME vector lies in the span of the uniform vector and
the power transfer distribution factor rows of the binding lines. Its rank
\(r\) is therefore at most one more than the number of binding/congested lines. This
structure makes \(r\) independent scalar observations necessary and sufficient
for exact recovery. Across ten systems with nonzero operating emissions, from
14 to 1,354 buses, \(r\) ranges from 2 to 15. For instance, on a 300 bus system,
24 dispatch simulations recover all 300 LMEs. We also derive an
emissions exceedance bound under uncertain demand. The bound separates
variation while the nominal active set remains unchanged, the probability of
an active set change, and estimation error. Numerical results show that its
usable forecast error range depends on local active set geometry.
\end{abstract}

\section{Introduction}

Locational marginal emissions (LMEs) measure the change in total operating
emissions caused by one additional MWh of energy consumed at each bus during a
dispatch interval. Prior work uses
these sensitivities to guide flexible demand, including data center load
shifting outside the system operator's dispatch
\citep{lindberg2020datacenters,lindberg2022geographic}. Other formulations place
carbon objectives inside dispatch
\citep{chen2023carbonaware,sun2023risklimiting,gu2022bridging}. However, changing the set of active constraints changes the LMEs, and deterministic
validity ranges can be small \citep{nilges2025validity}.

We make the following two contributions. First, we derive an intrinsic basis for the LME
vector, prove that its rank \(r\) is at most one more than the number of binding
line flow constraints, and show that \(r\) independent observations are necessary and
sufficient for recovery. Second, we combine a forecast error distribution with
the set of demand changes that preserve the nominal active set, which we call
the critical region. The resulting probability bound separates emissions
variation inside the region, region exit, and estimation error. Numerical
experiments test both results across networks up to 1,354 buses.

Existing methods for computing LMEs use direct differentiation
\citep{fuentesvalenzuela2024dynamic}, LMP mappings within critical regions
\citep{he2024lmp}, or finite differences at \textit{every} bus. We focus on LME recovery
when an external or legacy dispatch simulator can be evaluated but its equations
or solver derivatives cannot be accessed.

\section{Congestion Basis and Simulation Count}
\label{sec:structure}

Let \(d\in\mathbb R^n\) denote net demand in MW and
\(g\in\mathbb R^{n_g}\) generation in MW for a one hour dispatch interval,
\(\Gamma\) the generator incidence matrix, and \(H\) the power transfer
distribution factor (PTDF) matrix, whose entry \(H_{ji}\) gives the flow on
line \(j\) from an injection at bus \(i\). We study the strictly convex DC
dispatch
\begin{equation}
g^\star(d)=\arg\min_g\Big\{\tfrac12 g^\top Qg+a^\top g \;\Big|\;
\bone^\top(\Gamma g-d)=0,\;
|H(\Gamma g-d)|\leq\bar f,\;
\underline g\leq g\leq\bar g\Big\},
\label{eq:dispatch}
\end{equation}
where \(Q\succ0\) is the quadratic cost matrix. Total hourly operating
emissions are \(\cE(d)\triangleq c^\top g^\star(d)\) for generator emission
rates \(c\). Fix a nominal demand \(d_0\) with a regular active set. This means
that the active constraint rows are linearly independent and their inequality
multipliers are positive. The demands preserving this active set form its
critical region, within which \(g^\star(d)\) is affine. Let \(\mathcal J\)
index the binding lines and let
\(s_j\in\{-1,1\}\) denote the binding direction. The active equalities are
\begin{equation}
\bone^\top\Gamma g=\bone^\top d,\qquad
s_jH_j\Gamma g=\bar f_j+s_jH_jd\quad(j\in\mathcal J),
\qquad g_k=\underline{g}_k\ \text{or}\ g_k=\bar{g}_k
\label{eq:active}
\end{equation}
for each binding generator limit \(k\). After dependent rows are removed, let
\(K\) stack the coefficients multiplying \(g\) in Equation~\eqref{eq:active}
and write the right side as \(b+Dd\). The rows of \(D\) are
\(\bone^\top\), the signed PTDF rows \(s_jH_j\), and zeros for generator
limits. Differentiating the first-order optimality conditions gives
\begin{equation}
G\triangleq \partial g^\star/\partial d\big|_{d_0}
=Q^{-1}K^\top(KQ^{-1}K^\top)^{-1}D,\qquad
\ell\triangleq G^\top c ,
\label{eq:derivative}
\end{equation}
where \(\ell_i\) is the carbon emissions change per additional MW of demand sustained
for the one hour interval, with units of tonnes CO2/MWh. Equivalently,
\begin{equation}
\ell=D^\top(KQ^{-1}K^\top)^{-1}KQ^{-1}c .
\label{eq:lme-factor}
\end{equation}
Equation~\eqref{eq:lme-factor} shows that the cost and
generator data determine the coefficients multiplying \(D^\top\), but every
pattern across buses must come from a row of \(D\).

\begin{lemma}[Congestion basis for marginal emissions]
\label{prop:span}
Let \(q=|\mathcal J|\) be the number of binding line flow constraints at \(d_0\).
Then \(\ell\) lies in the range of \([\,\bone\;\;H_{\mathcal J}^\top\,]\), so
there are a scalar \(\alpha\) and a vector \(w\) of congestion weights with
\begin{equation}
\ell=\alpha\bone+H_{\mathcal J}^{\top}w,\qquad
r\triangleq \operatorname{rank}[\,\bone\;\;H_{\mathcal J}^\top\,]\leq q+1 .
\label{eq:span}
\end{equation}
Writing \(\Phi\) for any matrix whose \(r\) columns are an orthonormal basis of
that range and \(x\in\mathbb R^r\) for the coordinates of \(\ell\) in it gives
\(\ell=\Phi x\). We call \(r\) the \emph{congestion rank} of the system.
\end{lemma}

If there are no binding line flow limits, \(q=0\) and \(\ell=\alpha\bone\), so every bus has the same LME.
Each independent binding line can add at most one nodal pattern, namely its PTDF row.
Binding generator limits can change \(\alpha\) and \(w\), since they change the
dispatch response, but they add no pattern because their right sides do not
depend on demand. The result parallels the Laplacian subspace characterization
of admissible LMPs \citep{cheverezgonzalez2009admissible} and the energy and
congestion decomposition of LMPs \citep{he2024lmp}. The corresponding graph
view in Appendix~\ref{app:proofs} shows that the centered LME is harmonic away
from the endpoints of binding lines
\citep{baker2024locationalmarginalpricesobey}.

The lemma reduces recovery of all LME values from dispatch simulations to
estimating \(r\) coefficients rather than \(n\) bus values. For a demand step
\(h\) at bus \(i\), two simulations evaluate
\(d_0+h e_i\) and \(d_0-h e_i\). Their central difference equals \(\ell_i\)
when both demands remain in the same critical region. Since
\(\ell_i=\phi_i^\top x\), each observed bus supplies one linear equation in the
\(r\) unknown coefficients \(x\).

\begin{corollary}[Dispatch simulations for the LME vector]
\label{cor:recovery}
If \(r\) rows of \(\Phi\) form a nonsingular matrix, then observing \(\ell_i\)
at those \(r\) buses recovers the full vector exactly, so paired perturbations
there use \(2r\) dispatch simulations, or \(r+1\) with a cached base case,
against \(2n\) and \(n+1\) when every bus is perturbed. Fewer than
\(r\) scalar observations leave \(x\) underdetermined, since they impose fewer
than \(r\) independent linear constraints on it.
\end{corollary}

The lower bound in Corollary~\ref{cor:recovery} is within the declared
congestion span. It does not assert that finite differences are preferable to
direct derivatives. Any directional observation \(\ell^\top u\) gives the
equation \(x^\top(\Phi^\top u)\), and \(r\) independent directions also suffice.
For bus observations, pivoted QR factorization of \(\Phi^\top\) selects rows
that are independent and well conditioned. Thus, Corollary~\ref{cor:recovery}
separates the number and placement of bus perturbations. Congestion determines
how many are needed, while row conditioning determines where to perturb.

\section{Emissions Exceedance under Active Set Changes}
\label{sec:certificate}

Write realized demand as \(d=d_0+\xi\), where \(d_0\) is the forecast net demand
and \(\xi\) is its centered forecast error. For every direction
\(u\in\mathbb R^n\) and scalar \(\lambda\in\mathbb R\), assume
\(\mathbb E\exp(\lambda u^\top\xi)\leq\exp(\lambda^2u^\top\Sigma u/2)\),
so \(\xi\) is subgaussian with variance proxy
\(\Sigma\succeq0\), a matrix playing the role of a covariance that need not be
diagonal. Within the nominal critical region, the emissions change is
\(\ell^\top\xi\), whose variance proxy is
\begin{equation}
v=\ell^\top\Sigma\ell=x^\top Mx,\qquad
M=\Phi^\top\Sigma\Phi .
\label{eq:variance}
\end{equation}
Although \(\ell\) has \(n\) entries, its emissions uncertainty within the region depends only on the
\(r\) coefficients \(x\) and the compressed matrix \(M\in\mathbb R^{r\times r}\).
The same congestion dimension therefore controls both recovery and uncertainty
propagation while the active set remains fixed.

The nominal active constraints stay valid on the polytope
\(\cR=\{\xi\mid F_j^\top\xi\leq\tau_j,\ j=1,\ldots,p\}\). Here
\(F_j\in\mathbb R^n\) is the outward normal of facet \(j\) and \(\tau_j>0\) is
its slack at the forecast \(d_0\). The facets enforce inactive primal
constraints and nonnegative multipliers for active inequalities
(Appendix~\ref{app:region}). Applying a
subgaussian tail bound to each facet and a union bound gives the
region exit bound
\begin{equation}
\epscr=\min\!\left\{1,\sum_{j=1}^{p}
\exp\!\left[-\frac{\tau_j^2}{2F_j^\top\Sigma F_j}\right]\right\}.
\label{eq:switch}
\end{equation}
For Gaussian errors we replace each exponential with the exact facet
probability \(\Pr\{Z>\kappa_j\}\), where \(Z\) is standard normal and
\(\kappa_j=\tau_j/\sqrt{F_j^\top\Sigma F_j}\). We call
\(\kappa_{\min}=\min_j\kappa_j\) the active set margin. It is the smallest
standardized distance to a critical region facet, where either an inactive
primal constraint becomes binding or an active multiplier reaches zero. It is a useful summary, while the
complete bound in Equation~\eqref{eq:switch} retains every facet.

\begin{theorem}[Operating emissions exceedance bound]
\label{thm:certificate}
For any \(z>0\),
\begin{equation}
\Pr\{\cE(d_0+\xi)-\cE(d_0)\geq z\}
\leq \exp[-z^2/(2v)]+\epscr .
\label{eq:tail}
\end{equation}
\end{theorem}

The first term bounds the carbon emission increase while the nominal active
constraints hold, and Equation~\eqref{eq:switch} bounds the probability of
leaving that region. Keeping them separate shows which term limits the result.
For a target exceedance probability \(\delta\), Equation~\eqref{eq:tail} gives
a finite threshold only if \(\epscr<\delta\); otherwise region exit alone
reaches the target probability
\citep{nilges2025validity}. Figure~\ref{fig:certificate} shows both events.

\paragraph{Estimation from selected perturbations.}
A paired perturbation at bus \(i_t\) returns
\(y_t=[\cE_{\omega_t}(d_0+h e_{i_t})-\cE_{\omega_t}(d_0-h e_{i_t})]/2h
=\phi_{i_t}^\top x+\eta_t\),
where both perturbed demands must stay in the nominal critical region and
\(\eta_t\) is conditionally subgaussian and represents scenario sampling error
in a stochastic simulator. With ridge parameter \(\gamma>0\) and
\(V_t=\gamma I+\sum_{s\leq t}\phi_{i_s}\phi_{i_s}^\top\), a linear regression
confidence sequence under a declared \(\|x\|_2\leq\bar x\) bounds
\(\|\widehat x_t-x\|_{V_t}\) at every step simultaneously with
probability at least \(1-\delta_{\rm est}\) \citep{abbasi2011improved}, which
permits valid stopping based on the observations. Appendix~\ref{app:proofs}
turns this confidence set into a closed form \(\overline v_t\geq v\).
Our variance informed rule selects the next bus to maximize the reduction
\(\phi_i^\top V_t^{-1}MV_t^{-1}\phi_i/(1+\phi_i^\top V_t^{-1}\phi_i)\) in
\(\tr(MV_t^{-1})\), an $A$-optimal experimental design weighted
by \(M\), so it covers the directions carrying emissions variance first
\citep{fiez2019transductive,valko2014spectral}. Combining \(\overline v_t\)
with Theorem~\ref{thm:certificate}, and assuming future forecast error is
independent of simulation noise, gives the allowance
\(z_t=\sqrt{2\overline v_t\log(1/\delta_{\rm op})}\). Its total exceedance
probability is at most \(\delta_{\rm op}+\epscr+\delta_{\rm est}\) at any
stopping time.

\section{Numerical Results}
\label{sec:results}

We test 14 PGLib-OPF systems \citep{babaeinejadsarookolaee2019pglib} with
PGLib-CO2 rates \citep{cho2026pglibco2}. Ten have nonzero operating emissions;
the other four dispatch only generators with zero benchmark rates and have a
zero LME (Appendix~\ref{app:screening}). A transparent DC dispatch provides
exact derivatives for comparison (Appendix~\ref{app:repro}).

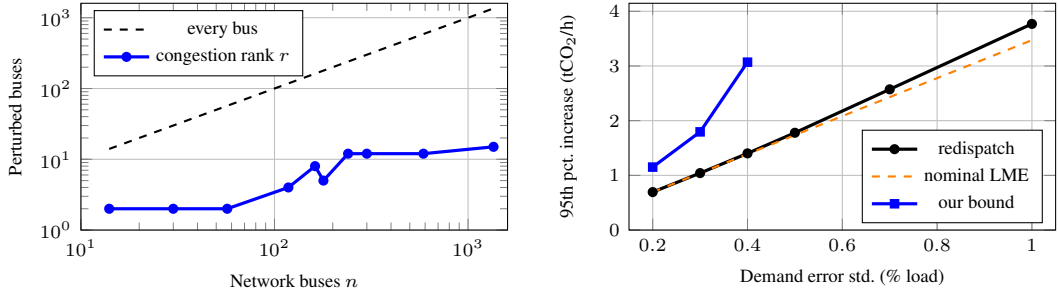
\begin{figure}[t]
  \vspace{-1mm}
  \centering
  \begin{tikzpicture}[
    every axis/.append style={
      scale only axis,
      width=0.404\textwidth,
      height=3cm,
      grid=major,
      tick label style={font=\scriptsize},
      label style={font=\scriptsize},
    },
  ]
    \begin{loglogaxis}[
      name=scalingplot,
      xlabel={Network buses \(n\)},
      ylabel={Perturbed buses},
      xmin=10,xmax=1600,
      ymin=1,ymax=1600,
      legend style={font=\scriptsize,at={(0.03,0.97)},anchor=north west},
    ]
      \addplot[thick,dashed,black]
        table[x=buses,y=full_coordinate_experiments,col sep=comma]
        {results/aggregate/scaling_plot.csv};
      \addlegendentry{every bus}
      \addplot[very thick,blue,mark=*,mark size=1.3]
        table[x=buses,y=experiments_for_exact_recovery,col sep=comma]
        {results/aggregate/scaling_plot.csv};
      \addlegendentry{congestion rank \(r\)}
    \end{loglogaxis}
    \begin{axis}[
      at={(scalingplot.east)},
      anchor=west,
      xshift=0.115\textwidth,
      xlabel={Demand error std.\ (\% load)},
      ylabel={95th pct.\ increase (tCO\(_2\)/h)},
      ylabel style={font=\scriptsize,at={(axis description cs:-0.10,0.5)}},
      xmin=0.15,xmax=1.05,
      ymin=0,
      legend style={font=\scriptsize,at={(0.97,0.03)},anchor=south east},
      unbounded coords=jump,
    ]
      \addplot[very thick,black,mark=*,mark size=1.3]
        table[x=load_forecast_std_percent,y=empirical_quantile,col sep=comma]
        {results/aggregate/emissions_tail_curve.csv};
      \addlegendentry{redispatch}
      \addplot[thick,dashed,orange]
        table[x=load_forecast_std_percent,y=linear_quantile,col sep=comma]
        {results/aggregate/emissions_tail_curve.csv};
      \addlegendentry{nominal LME}
      \addplot[very thick,blue,mark=square*,mark size=1.2]
        table[x=load_forecast_std_percent,y=selected_query_threshold_median,col sep=comma]
        {results/aggregate/emissions_tail_curve.csv};
      \addlegendentry{our bound}
    \end{axis}
  \end{tikzpicture}
  \caption{%Congestion sets the simulation count, and the bound withdraws rather than degrades.
  The congestion rank \(r\), between 2 and 15 across ten systems, is much lower than the number of buses \(n\) (left). On the 57-bus system, the
  nominal LME vector underestimates the realized 95th percentile once redispatch
  begins, while our bound stays above it and is not reported beyond a \(0.4\%\)
  nodal demand forecast error standard deviation, where active set exit leaves
  no probability for the within region tail (right).}
  \label{fig:structure}
  \vspace{-3mm}
\end{figure}

\paragraph{Simulation count.}
Figure~\ref{fig:structure} shows that the number of perturbed buses tracks
congestion and not network size. Across 14 to 1,354 buses, the rank \(r\) ranges
from 2 to 15, so the reduction \(n/r\) grows with size and reaches 90 on the
largest. On the 300 bus system, 12 selected central
differences recover the LME vector to \(1.2\times10^{-5}\) relative error using
24 simulations rather than 600 (Table~\ref{tab:full}). Dependent PTDF rows can
make \(r<q+1\), as on the 240 bus system where \(q=13\) and \(r=12\).
Omitting a binding line biases recovery by up to \(44\%\), whereas extra lines
only increase the simulation count. A residual from observations beyond the
declared rank detects this error (Appendix~\ref{app:details}). Under simulation
noise, the variance informed rule bounds emissions variance to within \(18\%\)
after 20 simulations, before the LME vector is identified
(Figure~\ref{fig:design}).

\paragraph{Emissions exceedance after redispatch.}
Figure~\ref{fig:structure} and Table~\ref{tab:tail} stress the 57-bus system
with 5,000 shared Gaussian demand errors per scale. At a \(1\%\) nodal demand
forecast error scale,
\(23.1\%\) of draws leave the region, the nominal LME prediction underestimates
the redispatch 95th percentile increase of \(3.77\) tonnes CO2/h by \(8.6\%\),
and every threshold violation follows an active set change. The threshold
stays available through \(0.4\%\) nodal demand forecast error, where Appendix~\ref{app:details} attributes
its factor of \(2.2\) over the empirical quantile mostly to the subgaussian
tail bound. The distributional assumption, not the facet union, causes the
loosest exit bounds (Table~\ref{tab:exit}).

\paragraph{When the bound is available.}
With \(0.5\%\) reserved for estimation, \(\epscr<0.045\) up to nodal demand
forecast error scales from \(0.12\%\) to beyond \(4\%\) of load, with no
relation to network size
(Table~\ref{tab:geometry}). Network size therefore does not indicate whether an
LME remains informative under forecast uncertainty; the active set geometry
must be checked at the operating point. Correlation narrows this range when
\(a_\star>1\) and widens it when \(a_\star<1\). Here \(a_\star\) compares that
facet's squared response to a common demand error with the sum of its squared
responses to independent nodal errors (Appendix~\ref{app:details}). Nine of ten
systems narrow, the 588 bus system has \(a_\star=0.35\) and widens.

\paragraph{Conclusion.}
Small congestion rank can make recovery inexpensive on a large network, but proximity to a critical region
boundary can make the recovered vector unreliable. A reported LME vector should be accompanied
by its local validity. Operational use also requires models of timing,
information exchange, and participant response. Future work includes sharper region
exit bounds, extensions to multiperiod or nonconvex dispatch, and applying these results to other sensitivity structures such as locational marginal burden~\citep{talkington2024burden}.

\clearpage
\bibliographystyle{plainnat}
\bibliography{references}

\clearpage
\appendix

\begin{figure}[t]
  \centering
  \begin{tikzpicture}[
    x=1.12cm,
    y=1.12cm,
    line cap=round,
    line join=round,
    every node/.append style={font=\footnotesize},
  ]
    \def\crpath{
      (-2.35,-0.40) -- (-1.55, 1.10) -- ( 0.25, 1.50) --
      ( 1.95, 0.80) -- ( 2.35,-0.70) -- ( 0.55,-1.50) --
      (-1.65,-1.30) -- cycle
    }

    % Forecast error plane: outside the critical region the active set moves.
    \fill[orange!14,rounded corners=2pt] (-2.62,-1.75) rectangle (2.62,1.75);
    \fill[white] \crpath;
    \begin{scope}
      \clip \crpath;
      \fill[blue!16] (0.92,-1.80) rectangle (2.70,1.80);
    \end{scope}
    \draw[black!55,semithick] \crpath;
    \draw[orange!75!black,semithick] (0.55,-1.50) -- (-1.65,-1.30);

    % The halfspace runs across the plane; only its part in R is shaded.
    \draw[blue!60!black] (0.92,-1.336) -- (0.92,1.224);
    \draw[blue!30,densely dashed] (0.92,1.224) -- (0.92,1.70);
    \draw[blue!30,densely dashed] (0.92,-1.336) -- (0.92,-1.70);

    % Forecast error spread at the operating point.
    \begin{scope}[rotate around={15:(0,0)}]
      \draw[black!38] (0,0) ellipse [x radius=0.80,y radius=0.44];
      \draw[black!45,densely dashed] (0,0) ellipse [x radius=1.35,y radius=0.75];
    \end{scope}
    \draw[orange!75!black,densely dotted] (0,0) -- (-0.131,-1.438);
    \fill[black!80] (0,0) circle (1.05pt);
    \draw[-{Latex[length=1.7mm]},black!80,semithick] (0.06,0) -- (0.82,0);

    \node[anchor=north east,inner sep=1pt] at (-0.02,-0.05) {\(0\)};
    \node[fill=white,inner sep=0.8pt,anchor=south] at (0.46,0.04) {\(\ell\)};
    \node[text=black!60] at (-1.06,0.72) {\(\Sigma\)};
    \node[text=black!70,font=\small] at (-1.52,-0.82) {\(\cR\)};
    \node[anchor=north west,text=orange!70!black] at (-2.50,1.66)
      {\(\xi\notin\cR\)};
    \node[text=orange!75!black,inner sep=1pt,anchor=west] at (-0.10,-1.02)
      {\(\tau\)};

    % The tail event is called out from the margin.
    \node[font=\small,text=blue!65!black] (tailtag)
      at (5.12,1.10) {\(\ell^\top\xi\geq z\)};
    \draw[blue!45,densely dashed,-{Latex[length=1.6mm]}]
      (tailtag.west) -- (1.68,0.26);

    % The two shaded events are the two terms of the bound.
    \node[anchor=west,font=\normalsize,inner sep=0,align=left] at (3.00,-0.22) {%
      \(\Pr\{\Delta\cE\geq z\}\ \leq\
        \textcolor{blue!65!black}{e^{-z^2/(2v)}}
        \,+\,
        \textcolor{orange!70!black}{\epscr}\)\\[8pt]
      \(\phantom{\Pr\{\Delta\cE\geq z\}\ \leq\ }\)%
      {\small\textcolor{black!55}{\(v=\ell^\top\Sigma\ell\)}}};
  \end{tikzpicture}
  \caption{Geometry of Theorem~\ref{thm:certificate} in forecast error space,
  with \(\Delta\cE=\cE(d_0+\xi)-\cE(d_0)\) and level sets of \(\Sigma\). Inside the critical region \(\cR\) the active set is fixed
  and \(\Delta\cE=\ell^\top\xi\), so reaching \(z\) means entering the blue
  set, i.e. the halfspace inside \(\cR\). Every facet contributes to \(\epscr\), dominated by the facet closest to the nominal operating point, at slack \(\tau\).}
  \label{fig:certificate}
\end{figure}
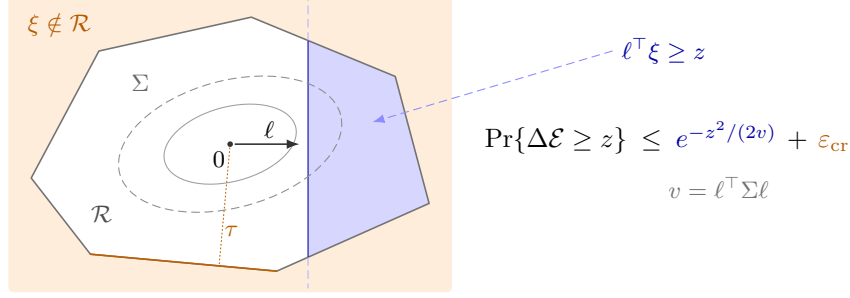

\section{Proofs}
\label{app:proofs}

\subsection{Congestion Basis for Marginal Emissions}

For a fixed active set, stationarity and feasibility are
\[
Qg+a+K^\top\nu=0,\qquad Kg=b+Dd .
\]
Here \(\nu\) collects the multipliers for the active equalities.
Eliminating \(\nu\) gives Equation~\eqref{eq:derivative}, and
\[
\ell
=D^\top(KQ^{-1}K^\top)^{-1}KQ^{-1}c .
\]
The power balance equality contributes \(\bone\). An active upper or lower line
constraint contributes a signed row of \(H_{\mathcal J}\). Generator bound
rows contribute zero because their right sides do not depend on demand. This
proves Equation~\eqref{eq:span}.

\paragraph{Graph interpretation.}
Let \(A\) denote the branch bus incidence matrix, \(B\) the diagonal matrix of
branch susceptances, and \(L=A^\top BA\) the weighted network Laplacian of a
connected network. In a shuntless network without phase shifters, the DC power flow relation gives
\(H=BAL^\dagger\), where \(L^\dagger\) is the pseudoinverse of \(L\)
\citep{cheverezgonzalez2009admissible,baker2024locationalmarginalpricesobey}.
Substituting this identity into
Equation~\eqref{eq:span} gives
\[
\ell-\alpha\bone
=L^\dagger A_{\mathcal J}^\top B_{\mathcal J}w
\implies
L(\ell-\alpha\bone)=A_{\mathcal J}^{\top}B_{\mathcal J}w ,
\]
because the source on the right sums to zero. This is a discrete Poisson
equation whose source is supported at no more than
\(2q\) endpoints of the binding lines. We use \emph{harmonic} in the graph
theoretic sense. A bus value vector \(z\) is harmonic at bus \(i\) when
\((Lz)_i=0\), which means that \(z_i\) equals the susceptance weighted average
of its neighboring values \citep{chung1997spectral,shuman2013graph}.

Let \(\mathcal T\) collect the binding line endpoints. When \(q>0\), every
component remaining after \(\mathcal T\) is removed is adjacent to
\(\mathcal T\). If two vectors in the congestion span agree on \(\mathcal T\),
their difference is zero there and harmonic at every other bus. Uniqueness of
the discrete Dirichlet problem therefore makes the difference zero everywhere
\citep{chung1997spectral}. Thus, the values on \(\mathcal T\) determine the full
vector. If \(q=0\), the vector is uniform and one bus value suffices. Independent
row selection reduces these boundary observations to the minimum \(r\).

\subsection{Recovery from Selected Buses}

Let \(I\) contain \(r\) buses such that \(\Phi_I\) is nonsingular. Their
observations satisfy \(\ell_I=\Phi_Ix\), hence
\(x=\Phi_I^{-1}\ell_I\) and \(\ell=\Phi\Phi_I^{-1}\ell_I\). This proves
Corollary~\ref{cor:recovery}. Without the congestion structure, fewer than \(n\) sparse
observations cannot distinguish a zero vector from a marginal emissions vector
supported at an unobserved bus.

\subsection{Exceedance Bound under Active Set Changes}

On the event \(\{\xi\in\cR\}\) the dispatch stays affine because the active
constraints are fixed, so
\(\cE(d_0+\xi)-\cE(d_0)=\ell^\top\xi\) and
\[
\Pr\{\cE(d_0+\xi)-\cE(d_0)\geq z\}
\leq \Pr\{\ell^\top\xi\geq z\}+\Pr\{\xi\notin\cR\} .
\]
The Chernoff bound gives the first term in Equation~\eqref{eq:tail}. Applying the
same upper tail bound to every event \(\{F_j^\top\xi>\tau_j\}\) and then taking a
union bound gives Equation~\eqref{eq:switch}. This proves
Theorem~\ref{thm:certificate}.

\subsection{Confidence Bound for Emissions Variance}

Let \(e=x-\widehat x_t\). Expanding the variance proxy gives
\[
 x^\top Mx
=\widehat x_t^\top M\widehat x_t
+2e^\top M\widehat x_t+e^\top Me .
\]
The confidence event implies \(\|V_t^{1/2}e\|_2\leq\beta_t\). The Cauchy--Schwarz inequality bounds the cross term and a Rayleigh quotient bounds the
quadratic term, so
\begin{equation}
\overline v_t=
\widehat x_t^\top M\widehat x_t+
2\beta_t\|V_t^{-1/2}M\widehat x_t\|_2+
\beta_t^2\lambda_{\max}(V_t^{-1/2}MV_t^{-1/2})\geq v .
\label{eq:ucb}
\end{equation}
The confidence event holds simultaneously at every step, so the result also
holds at a stopping time selected from the data.

\section{Critical Region Construction}
\label{app:region}

Write an inactive inequality as \(A_jg\leq b_j+E_jd\). With \(g=g_0+G\xi\),
primal feasibility becomes
\[
(A_jG-E_j)\xi\leq b_j+E_jd_0-A_jg_0 .
\]
Thus this constraint contributes facet normal
\(F_j^\top=A_jG-E_j\) and slack
\(\tau_j=b_j+E_jd_0-A_jg_0\). For an active inequality, let
\(\mu_i(\xi)=\mu_{i,0}+N_i\xi\) denote its KKT multiplier. Dual feasibility
becomes
\[
(-N_i)\xi\leq\mu_{i,0},\quad
N=-(KQ^{-1}K^\top)^{-1}D .
\]
It contributes facet normal \(F_j^\top=-N_i\) and slack
\(\tau_j=\mu_{i,0}\). The regularity assumption makes every retained slack
positive at \(d_0\).
The implementation removes zero normals and dependent active rows, and it
converts demand directions from per unit to MW before evaluating
Equation~\eqref{eq:switch}.

\section{Full Screening Results}
\label{app:screening}

\begin{table}[H]
\centering
\setlength{\tabcolsep}{2.5pt}
\caption{Congestion sets the rank on every retained system, and the reduction
\(n/r\) improves with system size. Here \(q\) counts binding line constraints,
\(r\) is the congestion basis rank, \(E_{\rm span}\) is the span residual, and
\(E_{\rm P}\) is the graph Poisson residual.}
\label{tab:full}
\begin{tabular}{lrrrrrr}
\toprule
System & $n$ & $q$ & $r$ & $n/r$ & $E_{\mathrm{span}}$ & $E_{\mathrm{P}}$ \\
\midrule
14 & 14 & 1 & 2 & 7.0 & 1.4e-15 & 3.5e-14 \\
30 & 30 & 1 & 2 & 15.0 & 3.1e-15 & 8.1e-14 \\
57 & 57 & 1 & 2 & 28.5 & 4.3e-14 & 2.7e-13 \\
118 & 118 & 3 & 4 & 29.5 & 4.9e-15 & 2.9e-14 \\
162 & 162 & 7 & 8 & 20.2 & 8.6e-15 & 1.3e-14 \\
179 & 179 & 4 & 5 & 35.8 & 9.6e-15 & 3.6e-14 \\
240 & 240 & 13 & 12 & 20.0 & 2.7e-14 & 1.6e-13 \\
300 & 300 & 11 & 12 & 25.0 & 1.1e-13 & 3.6e-13 \\
588 & 588 & 11 & 12 & 49.0 & 1.1e-13 & 2.0e-11 \\
1354 & 1354 & 14 & 15 & 90.3 & 2.1e-14 & 2.6e-14 \\
\bottomrule
\end{tabular}

\end{table}

\begin{table}[H]
\centering
\caption{Four screened systems carry no operating emissions under the direct
emission boundary, so their LME vector is zero and the study does not apply.}
\label{tab:excluded}
\begin{tabular}{ll}
\toprule
System & exclusion reason \\
\midrule
73 & zero operating emissions under direct factors \\
197 & zero operating emissions under direct factors \\
500 & zero operating emissions under direct factors \\
793 & zero operating emissions under direct factors \\
\bottomrule
\end{tabular}

\end{table}

At the selected operating point, the four systems of Table~\ref{tab:excluded}
dispatch only generators that the PGLib-CO2 lookup assigns a zero direct
operating rate. All four also miss our declared sensitivity tolerance of
\(10^{-8}\), with relative disagreements between the analytical sensitivity and
the numerical active set derivative running from \(8\times10^{-8}\) to \(0.42\), and three of the four have an optimality system with condition number above
\(10^{9}\). The exception is the 197 bus system, whose optimality system is
well conditioned but whose disagreement is the largest of the four. We read
both patterns as degeneracy of the active set at those operating points, and we
exclude the systems on the emissions criterion, which can be checked before
computing any sensitivity.

\section{Additional Numerical Details}
\label{app:details}

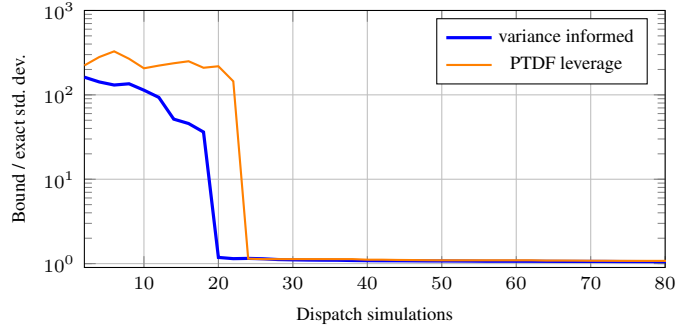
\begin{figure}[H]
  \centering
  \begin{tikzpicture}
    \begin{semilogyaxis}[
      scale only axis,
      width=0.55\textwidth,
      height=3.4cm,
      xlabel={Dispatch simulations},
      ylabel={Bound / exact std.\ dev.},
      xmin=2,xmax=80,
      ymin=0.9,ymax=1e3,
      grid=major,
      legend style={font=\scriptsize,at={(0.97,0.97)},anchor=north east},
      tick label style={font=\scriptsize},
      label style={font=\scriptsize},
    ]
      \addplot[very thick,blue]
        table[x expr=2*\thisrow{t},y=median_tail_margin_ratio,col sep=comma]
        {results/aggregate/experiment_risk_trace.csv};
      \addlegendentry{variance informed}
      \addplot[thick,orange]
        table[x expr=2*\thisrow{t},y=median_tail_margin_ratio,col sep=comma]
        {results/aggregate/experiment_leverage.csv};
      \addlegendentry{PTDF leverage}
    \end{semilogyaxis}
  \end{tikzpicture}
  \caption{Weighting the design by \(M\) bounds the emissions variance before
  the LME vector is identified. On the 300 bus system, where \(r=12\), variance informed selection reaches a usable bound after 20 dispatch simulations,
  while leverage selection must cover the full basis first and overestimates the
  standard deviation by a factor of 218 at that point. Once both cover the
  basis the advantage shrinks to \(2.6\%\), with interquantile ranges of
  \([1.079,1.091]\) and \([1.102,1.122]\) that do not overlap.}
  \label{fig:design}
\end{figure}

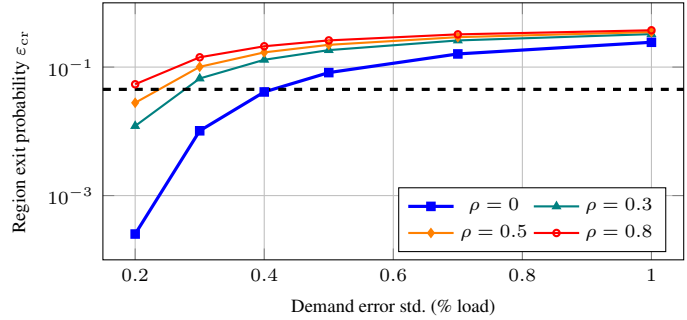
\begin{figure}[H]
  \centering
  \begin{tikzpicture}
    \begin{semilogyaxis}[
      scale only axis,
      width=0.55\textwidth,
      height=3.4cm,
      xlabel={Demand error std.\ (\% load)},
      ylabel={Region exit probability \(\epscr\)},
      xmin=0.15,xmax=1.05,
      ymin=1e-4,ymax=1,
      grid=major,
      legend style={
        font=\scriptsize,
        at={(0.97,0.03)},
        anchor=south east,
        legend columns=2,
      },
      tick label style={font=\scriptsize},
      label style={font=\scriptsize},
    ]
      \addplot[very thick,blue,mark=square*,mark size=1.2]
        table[x=load_forecast_std_percent,y=exit_bound,col sep=comma]
        {results/aggregate/correlated_tail_0p0.csv};
      \addlegendentry{\(\rho=0\)}
      \addplot[thick,teal,mark=triangle*,mark size=1.4]
        table[x=load_forecast_std_percent,y=exit_bound,col sep=comma]
        {results/aggregate/correlated_tail_0p3.csv};
      \addlegendentry{\(\rho=0.3\)}
      \addplot[thick,orange,mark=diamond*,mark size=1.4]
        table[x=load_forecast_std_percent,y=exit_bound,col sep=comma]
        {results/aggregate/correlated_tail_0p5.csv};
      \addlegendentry{\(\rho=0.5\)}
      \addplot[thick,red,mark=o,mark size=1.2]
        table[x=load_forecast_std_percent,y=exit_bound,col sep=comma]
        {results/aggregate/correlated_tail_0p8.csv};
      \addlegendentry{\(\rho=0.8\)}
      \addplot[very thick,dashed,black,forget plot]
        coordinates {(0.15,0.045) (1.05,0.045)};
    \end{semilogyaxis}
  \end{tikzpicture}
  \caption{On the 57-bus system, the nearest facet has aggregate response ratio
  \(a_\star=5.6\), so correlation costs margin and the exit probability crosses
  the remaining exceedance allowance, dashed, at a smaller demand error scale as \(\rho\)
  grows. The crossing moves from \(0.41\%\) at \(\rho=0\) to \(0.23\%\) at
  \(\rho=0.5\). Table~\ref{tab:geometry} reports \(a_\star\) for every system,
  including the one where it is below one and the ordering reverses.}
  \label{fig:exit}
\end{figure}

This appendix collects the supporting studies for Section~\ref{sec:results},
namely the perturbation step sweep in Table~\ref{tab:step}, the misdeclared
binding set in Table~\ref{tab:binding}, the congestion sweep in
Table~\ref{tab:sweep}, the tightness of the exit bound in Table~\ref{tab:exit},
the region geometry across systems in Table~\ref{tab:geometry}, the correlated
forecast error in Figure~\ref{fig:exit} and Table~\ref{tab:correlated}, the
ridge sensitivity in Table~\ref{tab:ridge}, and the design comparison in
Figure~\ref{fig:design}.

Write the correlated proxy as
\(\Sigma_\rho=\rho\,\sigma\sigma^\top+(1-\rho)\operatorname{diag}(\sigma^2)\),
which leaves every marginal standard deviation at \(\sigma_i\). Then
\(F_j^\top\Sigma_\rho F_j=(1-\rho+\rho a_j)\sum_i(\sigma_iF_{ji})^2\) with
\(a_j=(\sigma^\top F_j)^2/\sum_i(\sigma_iF_{ji})^2\), so the standardized
distance obeys \(\kappa_j(\rho)=\kappa_j(0)/\sqrt{1-\rho+\rho a_j}\) exactly.
Cauchy Schwarz gives \(a_j\in[0,n]\), with \(a_j=n\) when the facet responds
equally to every bus and \(a_j=0\) when it is a pure differential, and
correlation costs margin whenever \(a_j>1\). Table~\ref{tab:geometry} reports
the ratio at the nearest facet of each system, where the measured
\(\kappa_\star(0.5)\) matches this prediction to every digit shown.

The operating point search tests load scales \(1.0,1.1,1.2\) and line limit
scales \(1.0,0.9,\ldots,0.3\) in fixed order. We add \(10^{-4}\) to every
quadratic cost coefficient so that the local derivative is unique. The base
uncertainty proxy for every system draws independent Gaussian nodal demand forecast errors with
standard deviation equal to one percent of nominal demand and a
\SI{0.25}{MW} floor. The stress test in Section~\ref{sec:results} multiplies
that proxy by scales from \(0.2\) to \(1.0\), so the reported demand error
standard deviation of \(0.2\%\) to \(1.0\%\) of load is the product of the two.

\begin{table}[H]
\centering
\caption{The perturbation step must be small enough to keep both perturbed
demands inside the critical region. A \SI{1}{MW} step recovers the vector to
\(1.2\times10^{-5}\), \SI{10}{MW} leaves the region and costs \(4.2\%\), and
\SI{100}{MW} makes the dispatch infeasible. Facet use is the largest fraction
of a facet slack consumed by any selected perturbation, and a value above one
predicts the loss of accuracy.}
\label{tab:step}
\begin{tabular}{rlrr}
\toprule
step $h$ (MW) & inside region & facet use & recovery error \\
\midrule
0.1 & yes & 0.08 & 2.2e-04 \\
1.0 & yes & 0.77 & 1.2e-05 \\
10.0 & no & 7.75 & 4.2e-02 \\
50.0 & no & 38.73 & 3.0e-01 \\
100.0 & no & 77.46 & infeasible \\
250.0 & no & 193.65 & infeasible \\
500.0 & no & 387.30 & infeasible \\
\bottomrule
\end{tabular}

\end{table}

\begin{table}[H]
\centering
\caption{A declared binding set that omits a binding line biases recovery,
while one that names extra lines costs simulations but stays exact. Errors are
relative and taken over all single line omissions on the 300 bus system.}
\label{tab:binding}
\begin{tabular}{lrlrr}
\toprule
declared binding set & variants & rank $r$ & median error & worst error \\
\midrule
exact & 1 & 12 & 1.2e-13 & 1.2e-13 \\
omit one binding line & 11 & 11 & 3.4e-02 & 4.4e-01 \\
add 2 slack lines & 1 & 14 & 1.1e-13 & 1.1e-13 \\
add 5 slack lines & 1 & 17 & 1.0e-13 & 1.0e-13 \\
\bottomrule
\end{tabular}

\end{table}

\begin{table}[H]
\centering
\caption{Congestion sweep on the 300 bus system. The number of binding lines
and the rank both depend on the operating point.}
\label{tab:sweep}
\begin{tabular}{rrrr}
\toprule
line limit multiplier & binding lines $q$ & rank $r$ & $n/r$ \\
\midrule
1.3 & 5 & 6 & 50.0 \\
1.2 & 6 & 7 & 42.9 \\
1.1 & 8 & 9 & 33.3 \\
1.0 & 11 & 12 & 25.0 \\
0.9 & 10 & 11 & 27.3 \\
\bottomrule
\end{tabular}

\end{table}

\begin{table}[H]
\centering
\caption{The Gaussian column sits within Monte Carlo error of the empirical frequency on every tested system, while the distribution free subgaussian column overshoots by up to a factor of five. Concentration is the share of the Gaussian bound carried by the single nearest facet, and it exceeds \(0.93\) on eight of ten systems. Frequencies use 2,000 Gaussian draws.}
\label{tab:exit}
\begin{tabular}{lrrrrrr}
\toprule
System & facets & $\kappa_{\min}$ & sub-G. & Gauss. & exit freq. & conc. \\
\midrule
14 & 46 & 7.86 & 0.000 & 0.000 & 0.000 & 1.00 \\
30 & 89 & 26.65 & 0.000 & 0.000 & 0.000 & 1.00 \\
57 & 168 & 0.70 & 0.785 & 0.243 & 0.236 & 1.00 \\
118 & 422 & 1.55 & 0.436 & 0.083 & 0.080 & 0.73 \\
162 & 581 & 3.12 & 0.008 & 0.001 & 0.000 & 1.00 \\
179 & 556 & 1.12 & 0.596 & 0.141 & 0.143 & 0.94 \\
240 & 1034 & 1.44 & 0.908 & 0.176 & 0.166 & 0.43 \\
300 & 891 & 0.79 & 0.731 & 0.215 & 0.222 & 1.00 \\
588 & 1468 & 0.21 & 1.000 & 0.420 & -- & 0.99 \\
1354 & 4239 & 2.36 & 0.062 & 0.009 & -- & 1.00 \\
\bottomrule
\end{tabular}

\end{table}

\begin{table}[H]
\centering
\caption{The bound is available across the size range, and the aggregate
response ratio \(a_\star\) of the nearest facet predicts the sign of the
correlation effect exactly. Availability is the largest nodal demand forecast error standard
deviation, as a percentage of load, at which the exit term still leaves room in
the exceedance allowance. Systems with \(a_\star>1\) lose range under correlation and
the 588 bus system, with \(a_\star=0.35\), gains it. The facet type does not
predict the sign.}
\label{tab:geometry}
\begin{tabular}{lrrlrr}
\toprule
System & $\kappa_{\min}$ & $a_\star$ & nearest facet & avail.\ \(\rho{=}0\) & avail.\ \(\rho{=}0.5\) \\
\midrule
14 & 7.86 & 5.9 & generator & $>$4.0 & 2.50 \\
30 & 26.65 & 6.1 & generator & $>$4.0 & $>$4.0 \\
57 & 0.70 & 5.6 & generator & 0.41 & 0.23 \\
118 & 1.55 & 35.6 & generator & 0.86 & 0.21 \\
162 & 3.12 & 26.3 & line & 1.84 & 0.50 \\
179 & 1.12 & 17.1 & line & 0.66 & 0.22 \\
240 & 1.44 & 1.1 & generator & 0.72 & 0.41 \\
300 & 0.79 & 24.5 & generator & 0.47 & 0.13 \\
588 & 0.21 & 0.3 & generator & 0.12 & 0.15 \\
1354 & 2.36 & 61.6 & generator & 1.39 & 0.25 \\
\bottomrule
\end{tabular}

\end{table}

\begin{table}[H]
\centering
\caption{Correlated demand error narrows the usable range on the 57 bus system.
The demand error standard deviation at each bus is held fixed, so the only
change is the correlation \(\rho\) between buses. Thresholds are in tonnes CO2/h and a dash
means the exit term has consumed the exceedance allowance.}
\label{tab:correlated}
\begin{tabular}{rrrrrr}
\toprule
$\rho$ & demand error std. (\% load) & $\kappa_{\min}$ & exit freq. & $\varepsilon_{\rm cr}$ & threshold \\
\midrule
0.0 & 0.20 & 3.48 & 0.000 & 0.000 & 1.15 \\
0.0 & 0.30 & 2.32 & 0.007 & 0.010 & 1.80 \\
0.0 & 0.40 & 1.74 & 0.037 & 0.041 & 3.07 \\
0.0 & 0.50 & 1.39 & 0.073 & 0.082 & -- \\
0.0 & 0.70 & 0.99 & 0.152 & 0.160 & -- \\
0.0 & 1.00 & 0.70 & 0.231 & 0.243 & -- \\
0.3 & 0.20 & 2.25 & 0.010 & 0.012 & 2.17 \\
0.3 & 0.30 & 1.50 & 0.064 & 0.066 & -- \\
0.3 & 0.40 & 1.13 & 0.122 & 0.130 & -- \\
0.3 & 0.50 & 0.90 & 0.173 & 0.184 & -- \\
0.3 & 0.70 & 0.64 & 0.250 & 0.260 & -- \\
0.3 & 1.00 & 0.45 & 0.317 & 0.326 & -- \\
0.5 & 0.20 & 1.91 & 0.025 & 0.028 & 2.86 \\
0.5 & 0.30 & 1.28 & 0.100 & 0.101 & -- \\
0.5 & 0.40 & 0.96 & 0.167 & 0.169 & -- \\
0.5 & 0.50 & 0.77 & 0.220 & 0.222 & -- \\
0.5 & 0.70 & 0.55 & 0.286 & 0.292 & -- \\
0.5 & 1.00 & 0.38 & 0.346 & 0.351 & -- \\
0.8 & 0.20 & 1.61 & 0.053 & 0.054 & -- \\
0.8 & 0.30 & 1.07 & 0.144 & 0.142 & -- \\
0.8 & 0.40 & 0.80 & 0.209 & 0.211 & -- \\
0.8 & 0.50 & 0.64 & 0.261 & 0.260 & -- \\
0.8 & 0.70 & 0.46 & 0.323 & 0.323 & -- \\
0.8 & 1.00 & 0.32 & 0.374 & 0.374 & -- \\
\bottomrule
\end{tabular}

\end{table}

\begin{table}[H]
\centering
\caption{
The confidence radius grows like \(R\sqrt{r\log(1/\gamma)}\) through the log
determinant term and like \(\sqrt\gamma\,\bar x\) through the declared
norm bound, which places an interior optimum at the value used in the paper.
Values are medians over 100 trials after 40 dispatch simulations on the 57 bus
system.}
\label{tab:ridge}
\begin{tabular}{rrr}
\toprule
$\gamma$ & $\overline v_t/v$ & threshold inflation \\
\midrule
$10^{-14}$ & 1.155 & 1.075 \\
$10^{-12}$ & 1.144 & 1.070 \\
$10^{-10}$ & 1.134 & 1.065 \\
$10^{-8}$ & 1.125 & 1.061 \\
$10^{-6}$ & 1.143 & 1.069 \\
$10^{-4}$ & 1.472 & 1.213 \\
$10^{-2}$ & 7.369 & 2.715 \\
$10^{-1}$ & 32.603 & 5.710 \\
$10^{0}$ & 105.577 & 10.275 \\
\bottomrule
\end{tabular}

\end{table}

The recovery study on the 300 bus system uses a \SI{1}{MW} central difference,
and a pivoted QR factorization selects the perturbation buses. The noisy study
uses
\[
\beta_t=R\sqrt{
\log\!\frac{\det V_t}{\det(\gamma I)}
+2\log(1/\delta_{\rm est})}
+\sqrt{\gamma}\bar x
\]
with \(R=0.02\) tonnes CO2/MWh, which is \(2.9\%\) of the mean marginal
emission rate on that system and stands for the scenario sampling error a
stochastic simulator would carry, together with \(\gamma=10^{-8}\),
\(\delta_{\rm est}=0.005\), and
\(\bar x=5\sqrt n\). The realized \(\|x\|_2=12.5\) sits well below
\(\bar x=86.6\), and since \(\sqrt\gamma\bar x\approx10^{-3}\) here,
the declared bound contributes almost nothing to \(\beta_t\) and a deployment
would only need it to be finite. One noisy observation represents a pair of
load perturbations and therefore requires two dispatch simulations when the
base case is not cached. After 40 simulations, uniform bus selection has median
\(\sqrt{\overline v_t/v}=433\), because it leaves directions relevant to \(M\)
nearly unobserved. It is an ill conditioned diagnostic rather than evidence of
a comparable advantage over leverage selection.

The separate region diagnostic reuses 50,000 standard Gaussian draws across 18 forecast
error scales on the 118 bus system. The complete experiment on the 57 bus
system uses 5,000 draws at each scale and the Gaussian facet union bound.
Summary exit estimates use 2,000 draws on systems with up to 300 buses, and we
do not report an empirical frequency for larger systems.

\begin{table}[H]
\centering
\small
\setlength{\tabcolsep}{2.4pt}
\caption{Complete emissions experiment on the 57 bus system. Thresholds are in
tonnes CO2/h. The selected perturbation column is the median over 100 trials of
variance informed selection after 40 dispatch simulations, and it assigns
\(0.5\%\) allowed exceedance probability to estimation. A dash means the remaining \(5\%\)
exceedance allowance leaves no positive probability for the within region
tail. The solve count includes every realization outside the region and 100
validation draws inside it.}
\label{tab:tail}
\begin{tabular}{rrrrrrrr}
\toprule
\shortstack{demand forecast\\error std. (\% load)} & $\kappa_{\min}$ & exit freq. & $\varepsilon_{\rm cr}$ & empirical $q_{.95}$ & known $v$ $z_{.95}$ & selected perturb. $z_{.95}$ & solves \\
\midrule
0.20 & 3.48 & 0.000 & 0.000 & 0.69 & 1.07 & 1.15 & 100 \\
0.30 & 2.32 & 0.007 & 0.010 & 1.04 & 1.60 & 1.80 & 136 \\
0.40 & 1.74 & 0.037 & 0.041 & 1.40 & 2.13 & 3.07 & 283 \\
0.50 & 1.39 & 0.073 & 0.082 & 1.78 & 2.67 & -- & 466 \\
0.70 & 0.99 & 0.152 & 0.160 & 2.57 & 3.73 & -- & 860 \\
1.00 & 0.70 & 0.231 & 0.243 & 3.77 & 5.33 & -- & 1255 \\
\bottomrule
\end{tabular}

\end{table}

The reported threshold at a nodal demand forecast error standard deviation of \(0.4\%\) exceeds
the empirical 95th percentile by a factor of \(2.2\), which decomposes into
\(1.49\) for the subgaussian tail bound, \(1.25\) for splitting the exceedance
allowance with the exit term, and \(1.15\) for estimation after 40 simulations.
Under a Gaussian forecast error, the exact quantile replaces the Chernoff step
of Theorem~\ref{thm:certificate}, which isolates the price of the weaker
subgaussian assumption. With a known \(v\), that refinement gives \(1.43\)
tonnes CO2/h against an empirical \(1.40\), and it stays within \(3\%\) at
every scale where the answer is available. At \(0.7\%\) and above, where the
answer is not available, the same Gaussian threshold falls below the true
quantile, which is the bias that the exit term prevents.

\section{Reproducibility}
\label{app:repro}

The Julia implementation and numerical results are available at
\url{https://github.com/cameronkhanpour/LMEs-and-Carbon-Concentration}.
The repository includes experiment scripts,
benchmark data and provenance, generated CSV outputs, and automated tests.
The experiment manifest records package versions, random seeds, solver
settings, exclusion criteria, uncertainty parameters, and sample counts. The
README provides instructions for running a short validation case, the test
suite, and the complete numerical pipeline.

\end{document}